\documentclass[twocolumn,showpacs,pre,amsfonts,amsmath,amssymb]{revtex4}
\usepackage{graphicx}

\begin{document}

\title{
Finite-element analysis of contact between elastic self-affine surfaces
}

\author{S. Hyun, L. Pei, J.-F. Molinari, and M. O. Robbins}
\affiliation{Department of Physics \& Astronomy, Department of Mechanical Engineering,
Johns Hopkins University,
Baltimore, Maryland 21218, U.S.A}

\date{\today}

\begin{abstract}

Finite element methods are used to study non-adhesive,
frictionless contact between elastic solids
with self-affine surfaces.
We find that the
total contact area rises linearly with load at small loads.
The mean pressure in the contact regions is independent of load and
proportional to the rms slope of the surface.
The constant of proportionality is nearly independent of Poisson
ratio and roughness exponent and lies between previous analytic predictions.
The contact morphology is also analyzed.
Connected contact regions have a fractal area and perimeter.
The probability of finding a cluster of area $a_c$ drops as $a_c^{-\tau}$
where $\tau $ increases with decreasing roughness exponent.
The distribution of pressures shows an exponential tail that is
also found in many jammed systems.
These results are contrasted to simpler models and experiment.

\end{abstract}

\maketitle
\newpage 

\section{Introduction}
\label{sec: introduction}

The forces of friction and adhesion between two surfaces are determined
by the interactions between atoms at their interface.
Most surfaces are rough enough that atoms are only close enough to
interact strongly in areas where peaks or asperities
on opposing surfaces overlap.
Experiments \cite{bowden86,dieterich94,dieterich96,berthoud98}
and theory \cite{greenwood84,greenwood01,volmer97,persson01,persson02,batrouni02,ciavarella00,roux93}
show that this real area of contact $A$ is
often much smaller than the projected area $A_0$ of the surfaces.
They have also correlated \cite{dieterich94,dieterich96,bowden86} the increase
in friction with normal load $W$ to a corresponding increase in $A$.

Given the importance of $A$ it is not surprising that there
have been many theoretical studies of the factors that determine it
in different limits.
Some have examined the plastic limit where the local pressure
is large enough to flatten asperities \cite{bowden86}.
The mean pressure in the contacts is $W/A$ and in the
simplest model this has a constant value that is proportional
to the hardness.
The resulting linear relation between $W$ and $A$ is often
given as an explanation for the linear rise of friction with load
\cite{bowden86,johnson85}.

The behavior of elastic contacts is more complicated.
In the Hertzian limit where friction and adhesion are ignored,
the contact area between a sphere and a flat rises only
as $W^{2/3}$.
However, Greenwood and Williamson's pioneering work \cite{greenwood66}
showed that a nearly linear relation between $W$ and $A$ was
obtained if one considered a large number of asperities of
different height.
While this calculation assumed spherical asperities with
a uniform radius, a linear relation was also obtained
when it was extended by Bush {\it et al.} \cite{bush75}
to include both a distribution of radii
and aspherical asperities.
These calculations consider explicit probability distributions
for peaks and sum the Hertzian contact areas calculated for
each peak without including correlations between peaks.
Persson \cite{persson01} has recently presented a 
different approach, motivated by the fact that many surfaces
have roughness on all length scales that can be described
by self-affine scaling \cite{mandelbrot,krim95,meakin98}.
His calculation considers the scaling of stress and contact
area with the lengthscale to which surface features are resolved.
Nevertheless, his final result for $A$ only differs from the
earlier work of Bush {\it et al.} \cite{bush75}
by a constant factor of $\pi/2$.

All of the above theories treat correlations
between contacting regions approximately.
Models based on asperities ignore spatial correlations
in asperity heights \cite{greenwood66,bush75}. 
Persson's calculation includes height correlations
and becomes exact in the limit of complete contact \cite{persson02}.
However, when $A<A_0$,
correlations between local pressures and contacts are only
included in an average way.
As a result,
screening of small bumps on the side of larger bumps,
may not be included completely.
Such correlations could change $A$, and also
the distribution of pressure along the surface.
The range over which $A$ is proportional to load
is also difficult to determine from analytical theories.
Indeed, as discussed below, the work of Bush {\it et al.} \cite{bush75}
seems to suggest that the linear region is confined to infinitesimally
small loads for a self-affine surface.

A recent numerical study by Borri-Brunetto {\it et al.}
\cite{borri01}
calculated $A$ and the spatial distribution of contact areas
using a method that is restricted to the limit of zero Poisson ratio.
They found a substantial range where $A$ rose linearly with
load, but did not test analytic results for the slope specifically.
As in Persson's work \cite{persson01},
their focus was on the change in results with increasing resolution
of surface roughness.
Batrouni {\it et al.}\cite{batrouni02} have considered the scaling
of contact area with load using a similar method.
They find a slight deviation from linearity, but rule out
larger deviations predicted by previous scaling 
arguments \cite{roux93}.

In this paper, we present numerical calculations of contact area
and pressure distributions for a wide range of Poisson ratios,
loads, system sizes, self-affine scaling exponents and roughness
amplitudes.
We first show that $A$ has a well-defined thermodynamic limit,
that is for fixed small scale roughness the fraction of area
in contact at a given average normal pressure is independent of
system size.
The ratio of $A$ over load is shown to scale as the inverse
of the rms surface slope, with a coefficient that lies between
the results of Bush {\it et al.} \cite{bush75} and Persson \cite{persson01}.
The dependence of $A/W$ on roughness exponent and Poisson
ratio is also obtained.
This allows $A$ to be predicted for any elastic rough surface.

We next describe the contact morphology and
distribution of connected contact areas.
The probability of finding a connected region of area $a_c$
falls off as a power law, $P(a_c) \propto a_c^{-\tau}$,
where $\tau$ depends only on roughness exponent $H$.
For $H < 0.9$, $\tau$ is greater than 2, and the mean contact size is
always comparable to the resolution of the calculation.
As first noted by Greenwood and Williamson \cite{greenwood66},
the linear rise in $A$ with load reflects a linear
increase in the number of contacts without any increase in
their mean size or probability distribution.
Our results are contrasted to a common model where contacts
form in the regions where undeformed surfaces would overlap.
This approximation has been used to interpret optical images
of contacts \cite{dieterich94,dieterich96},
and by Greenwood and Wu as a method of estimating the statistics
of asperity sizes \cite{greenwood01}.
We show that it gives much too large a total contact area,
and a qualitatively different distribution of $a_c$
with $\tau < 2$.
Optical methods may include regions that are merely close
to touching as part of the contact.
We find that this can explain the observed discrepancy
between experiment and calculation.

The distribution of contact pressures $p$ is also studied.
We find that results for all system sizes, roughness amplitudes,
and roughness exponents collapse onto a universal curve
when the pressure is normalized by its mean value.
The mean value, $W/A$, can be obtained from the roughness
amplitude as described above.
The distribution decreases monotonically with increasing $p$,
and has an exponential tail at large $p$.
In contrast, approximate analytic results for the pressure
distribution decay as a Gaussian at large pressures \cite{persson01}.
Possible connections to exponential stress distributions
in jammed systems \cite{liu95,coppersmith96,liu01} are mentioned. 

In Sec. II, we describe the numerical procedures used to
generate self-affine surfaces, mesh them, and determine their
deformation using an explicit dynamic finite-element method.
In Section \ref{sec: result}, we present numerical results from the
contact analysis, including the contact area, contact morphology and
pressure distribution.
The final section presents conclusions from the current study
and discusses avenues for future research.

\section{Numerical Simulation}
\label{sec: simulation}

\subsection{Geometry}
\label{sec:geometry}

As in previous analytical and numerical calculations we use a well-known
result from contact mechanics to simplify the geometry.
If there is no friction or adhesion between two rough surfaces
and the surface slope is small, then elastic contact between them
can be mapped to contact between a single rough
surface and a rigid flat plane\cite{johnson85}.
The effective modulus $E'$ that controls the contact area
is given by
$1/ E' = (1-\nu_1^2)/E_1 + (1-\nu_2^2)/E_2$
where $\nu_i$ and $E_i$ are the Poisson ratios and Young's moduli
of the two surfaces. 
The height $h$ of the new rough surface is given by the difference between
the local heights of the original undeformed surfaces.
We consider the case where both of the surfaces have the same
self-affine scaling properties but are uncorrelated.
Then the height $h$ has the same scaling properties, but with
a larger amplitude.

Many surfaces have roughness on all length scales that can be described
by self-affine fractal scaling.
Unlike self-similar fractals, self-affine fractal surfaces exhibit different
scaling normal to the interface than along it.
We consider geometries with periodic boundary conditions in the $x-y$ plane
and specify the surface by the height $h$ along the $z$-axis.
For a self-affine surface, variations in the height over a lateral
length scale $\ell$ rise as $\ell^H$ where $H < 1$ is called 
the Hurst or roughness exponent.
Since $H<1$, the surface looks smoother at larger length scales.
Many researchers also specify the scaling by an effective fractal dimension
$d-H$, where $d$ is the spatial dimension.

\begin{figure}
\includegraphics[height=2.90in]{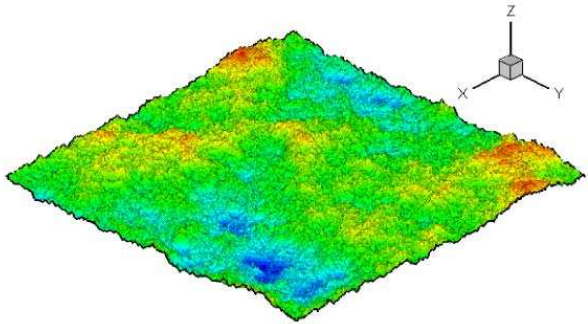}
\caption{A self-affine fractal surface image (256x256) generated by
the \textit{successive random midpoint} algorithm.
Heights are magnified
by a factor of 10 to make the roughness visible,
and the color varies from dark (blue) to light (red) with increasing height.}
\label{fig:surface}
\end{figure}

To generate three-dimensional self-affine fractal surfaces
with rms roughness $\Delta$ at small scales, we adopted the
successive random midpoint algorithm of Voss \cite{voss85,meakin98}.
The $x-y$ plane is divided into a uniform square grid with unit spacing
and $L$ nodes along each axis.
At the first step, the center of the entire grid is displaced by a height
chosen at random from a Gaussian distribution of width $\ell^H
\Delta$,
where $\ell = L/\sqrt{2}$ is the distance of the center from the corners. 
This center point then becomes one corner of new squares
that are rotated by 45$^{o}$ and have a new corner to center distance
$\ell$ that is smaller by a factor of $\sqrt{2}$.
The center of each new square is assigned a height equal to the average
of the corner heights plus a random number chosen from a Gaussian of width
$\ell^H \Delta$.
This process is iterated down to $\ell=1$, guaranteeing that the variation
in height scales with $\ell$ in the appropriate manner.
Figure \ref{fig:surface} shows a typical self-affine surface with $H=0.5$.
Note that the height variation is enhanced by a factor of ten
to make it visible in the figure.

It is common to test the scaling properties of self-affine surfaces
by calculating the Fourier transform $C(q)$
\begin{equation}
C(q) = (2\pi)^{-2} \int d^2 r \exp [-i {\bf q}\cdot {\bf r }] \ \ C({\bf r})
\label{eq:cq}
\end{equation}
of the height-height correlation function
\begin{equation}
C({\bf r }) =  < h({\bf r } +{\bf r '} )h({\bf r '})>
\label{eq:cr}
\end{equation}
where $\langle h({\bf r  }^\prime )\rangle =0$,
${\bf r}$ and ${\bf r}^\prime$ are vectors in the $x-y$ plane,
and the brackets indicate an average over ${\bf r}^\prime$.
The function $C(q)$ should be isotropic and decay as $q^{-2(1+H)}$.
We have generated surfaces with $H$ between 0.3 and 0.9
and verified that the corresponding $C(q)$ has the correct power law
scaling.
For a given $\Delta$, the values of $C(q)$ at large $q$ are very
insensitive to the random seed.
However, there are large fluctuations at small $q$ where the
roughness is dominated by the first few random numbers that are chosen
at the largest length scales.
These first random numbers also dominate the mean-squared
height variation:
\begin{equation}
C(r=0)= <|h({\bf r '})|^2> ,
\end{equation}
and we find large variations in $C(0)$ for surfaces 
with the same small
scale roughness $\Delta$.
As discussed below, the contact area is determined by $\Delta$ and
is relatively insensitive to fluctuations in $C(0)$.

\subsection{Mesh Generation}

\begin{figure}
\includegraphics[height=3.00in]{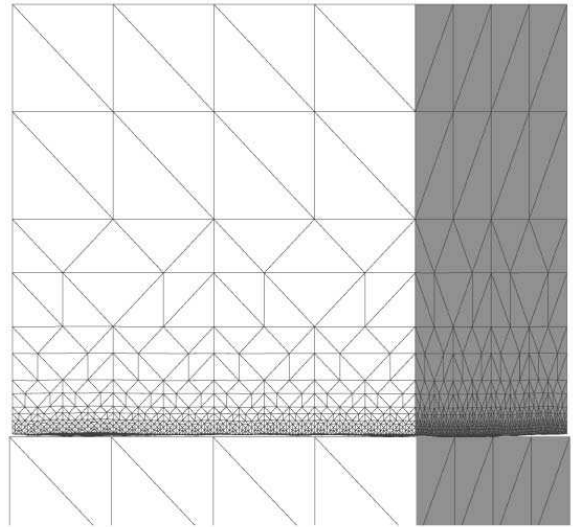}
\caption{Geometry of a finite element mesh in an elastic body (top)
with a self-affine surface that is pushed down on
a flat, rigid substrate.
Periodic boundary conditions are applied in the plane of the interface.}
\label{fig:FEM}
\end{figure}

As noted above, numerical simulations are done for a rough elastic surface
contacting a perfectly rigid flat surface.
A typical finite element mesh is illustrated in Figure \ref{fig:FEM}. 
The mesh is discretized with ten-node tetrahedral elements. These
elements contain three integration points and quadratically interpolate
the displacement field.
A coarse mesh is used for the rigid surface to improve numerical efficiency.
A fine mesh for the elastic surface is prepared in two stages.

First, a fine mesh for a flat surface is obtained using a longest edge
propagation path refinement scheme, which is ideally suited to obtain
strong mesh gradations as
well as preserving a high mesh quality \cite{molinari00c}. 
A cube of side $L=2^n$ is initially filled with a coarse mesh.
Each tetrahedral element at the outer surface is then divided
to produce twice as many surface nodes and the mesh is refined.
This process is repeated until surface nodes form a uniform square grid of
unit spacing.
Using this technique, meshes with $L$ up to $512$ are created.
The resulting grid contains $512\times 512$ surface nodes, about $911,000$
total nodes and $568,000$ elements.

Next, the desired surface heights $h(x,y)$ are imposed onto the contact surface.
Moving only the surface nodes produces badly distorted elements,
that would at best require impractically small time steps and
at worst produce negative Jacobians.
Thus all nodes are moved by a fraction of the local height that depends
on the initial height $z_0$ of the node above the bottom of the elastic cube
(Fig. \ref{fig:FEM}).
The magnitude of the change, $\Delta z$, decreases to zero at the top of the
cube so that the top surface remains flat.
The specific form for the displacement is
$\Delta z (x,y,z_0) = h(x,y)*(L-z_0)^a$ where $a= 6 $ usually gives good meshes.

\subsection{Finite Element Simulation}

The goal is to determine the equilibrium contact geometry at a given load.
An implicit approach is too memory intensive for the system sizes
of interest.
Instead we use an explicit integration algorithm combined with a dynamic
relaxation scheme.
Three different algorithms were compared to insure accuracy.
In the first, the top surface is given a small velocity and its impact
with the bottom surface is followed.
In the second, the displacement of the nodes at the top of the elastic cube
(Fig. \ref{fig:FEM}) is incremented at a fixed rate or in small discrete steps.
In the third, a constant force is applied to each of these nodes and
gradually incremented.
In the second and third algorithms kinetic energy is removed using the
method described below.
All three methods give equivalent results for the total area.
Unless otherwise noted,
the results presented below were obtained with the third algorithm.
We confirmed that the mean normal stress was independent of height to
ensure that stress had equilibrated throughout the system.

Within the Lagrangian framework, the finite-element discretization
of the field equations leads to a discrete system of equations:
\begin{equation} 
{\bf M}\ddot{{\bf x}}_{n+1} + {\bf F}_{n+1}^{int}({\bf x},\dot{{\bf x}}) = 
{\bf F}_{n+1}^{ext} 
\label{LinearMomentum} 
\end{equation} 
where ${\bf x}$ is the array of nodal coordinates, ${\bf M}$ the mass matrix,
${\bf F}_{n+1}^{ext}$ the external force array, and ${\bf F}_{n+1}^{int}$ the
internal force array arising from the current state of stress. 
The second-order accurate central difference scheme is used to discretize
Eq. (\ref{LinearMomentum}) in time \cite{belytschko83,hughes87}. 
A small time step was used in order to be below the stability limit
\cite{belytschko83}. 

The above equations conserve energy and will not converge to the
static equilibrium configuration.
Optimum convergence is achieved by removing a fraction of the kinetic energy
of each node at regular intervals.
The characteristic time for stress equilibration across the elastic
cube is given by the time $\tau_L$ for sound propagation across its height $L$.
Equilibrium is reached in a few $\tau_L$ by scaling all velocities
by a factor of 0.9 at intervals of $\tau_L/10$.
Other procedures gave equivalent results, but with longer run times.

The internal forces $F^{int}$ are calculated using a
linear elastic isotropic constitutive law. 
All our results are expressed in dimensionless form
by normalizing pressures by the effective modulus $E'$.
The Poisson ratio $\nu$ was varied from 0 to 0.45. 
Periodic boundary conditions are imposed at the contact surfaces to eliminate
boundary effects.

A contact algorithm is used only to enforce the impenetrability constraint
on the two surfaces.
Adhesive and frictional forces are not considered in the current work.
We adopt a conventional master/slave approach with a predictor/corrector split
within the Newmark time-stepping algorithm \cite{molinari01}. The
rough surface of the elastic cube is identified as slave while the
rigid surface is master. The predictor part of the Newmark algorithm neglects the contact constraints and,
therefore, consists of an unconstrained step, with the result: 
\begin{eqnarray} 
{\bf x}^{pred}_{n+1} &=& {\bf x}_n + \Delta t {\bf v}_n + \frac{{\Delta t}^2}{2} {\bf a}_n \\ 
{\bf v}^{pred}_{n+1} &=& {\bf v}_n + \frac{{\Delta t}}{2} {\bf a}_n 
\end{eqnarray} 
This predictor solution needs to be corrected in order to comply with the impenetrability constraints. 
The net result of imposing these constraints is a set of self-equilibrated contact forces that modify the predictor positions and velocities. Since the contact surfaces are presumed smooth, normals are well-defined and the surfaces can be unambiguously classified as master and slave. 
The final corrector configuration is therefore: 
\begin{eqnarray} 
{\bf x}^{S}_{n+1} &=& {\bf x}^{S,pred}_{n+1} - \frac{{\Delta t}^2}{2} 
\frac{{\bf N}^{S}_{n+1} + {\bf F}^{S}_{n+1}}{M^{S}} \\ 
{\bf v}^{S}_{n+1} &=& {\bf v}^{S,pred}_{n+1} - 
        {\Delta t} \frac{{\bf N}^{S}_{n+1} + {\bf F}^{S}_{n+1}}{M^{S}} 
\end{eqnarray} 
Here $M^{()}$ denotes the nodal mass, the superscript ${()}^S$
designates nodes that belong to the slave surface,
and
the vectors ${\bf N}$ and ${\bf F}$ are the normal and frictional
forces, respectively. 
Friction will not be taken into account in the remainder of the paper,
but is discussed in Ref. \cite{molinari01}. 
  
The formulation of an appropriate system of forces is obtained by considering
a configuration in which a master surface triangle
(facet of a tetrahedral finite element) is penetrated by several slave nodes. 
For each of the penetrating slave nodes, let $\delta$ be the normal depth of
penetration to be corrected by the contact forces. 
The contact constraints determine a local problem with the normal slave
force as an unknown,
which is obtained as a
direct function of the penetration $\delta$ and the master normal, ${\bf n}$: 
\begin{equation} 
{\bf N}^{S}_{n+1} = M^{S} \frac{2 \delta}{{\Delta t}^2} {\bf n}^{M}_{n+1} 
\label{slaves} 
\end{equation}

\subsection{Calculating the Contact Area}

The contact algorithm just described identifies all slave nodes on the top
surface that attempt to penetrate the flat bottom surface.
We obtain the total contact area by multiplying the number of penetrating
nodes $n_p$ by the area of the square associated with each node.
In most cases we report the fractional area $A/A_0=n_p/L^2$ where
$L^2$ is the total number of surface nodes.

\begin{figure}
\includegraphics[height=3.0in]{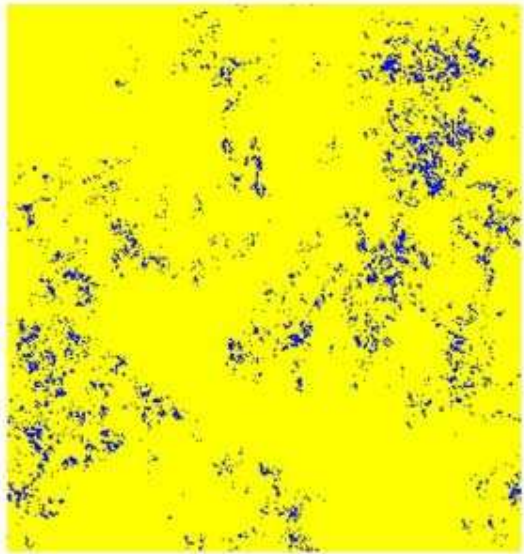}
\caption{
Regions of contact (dark) for a surface with $L=256$,
$H=1/2$, $\nu=0$, and $\Delta=0.082$.
The fraction of the area in contact $A/A_0 = 0.1$.
}
\label{fig:contact}
\end{figure}

In principle, the area associated with each node varies with
normal load if the Poisson ratio is non-zero.
However, tests using a Voronoi tessellation to determine the area
for each node gave equivalent results for the relatively smooth
surfaces considered here.
A Voronoi approach becomes important for rougher surfaces or
irregular grids.
A more fundamental ambiguity in the contact area comes from the fact
that contacts (Fig. \ref{fig:contact}) contain many disjointed regions
most of which contain only a few nodes.
The consequences of this are discussed in Sec. \ref{sec:area}.

There are also complications in defining the contact area in
experimental systems.
Dieterich and Kilgore's optical method identifies any region
where the surfaces are closer than some fraction of the wavelength
as in contact.
Due to the fractal nature of contacts (Fig. \ref{fig:contact})
this may overestimate the true area.
At the atomic scale, contact is difficult to define.
If one associates contact with a finite interaction, then $A/A_0$
would always be unity since the van der Waals interactions between
surfaces extend to arbitrary distances.
A more practical definition is to associate contact with the
separation at which the net interaction becomes strongly
repulsive due to the overlap of electrons on opposing surfaces.
This leads to a range of separations where surfaces are in contact.
As in the optical measurements, the contact area can be greatly
enhanced relative to the penetration definition used here.
Direct comparisons to atomistic models will be presented in another
paper \cite{Harrison}.

\section{Results}
\label{sec: result}

\subsection{Contact area vs external load}

\begin{figure}
\includegraphics[height=3.0in]{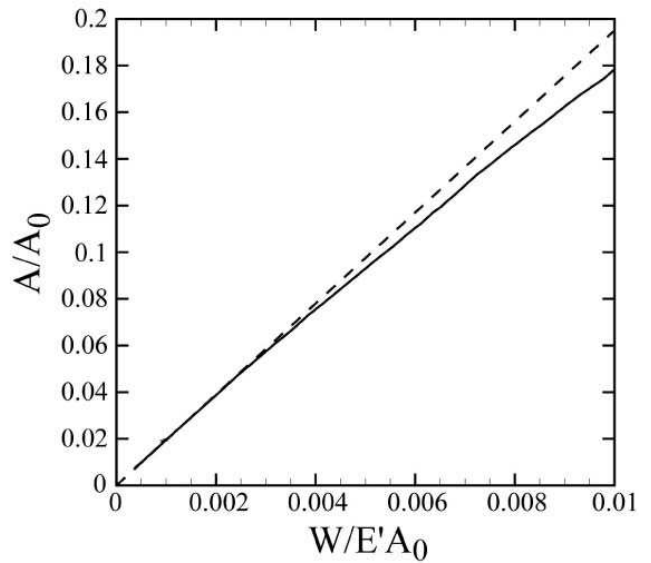}
\caption{
Fractional contact area $A/A_0$ (solid line)
as a function of the normalized load $ W/E' A_0$ 
for $L=256$, $\Delta = 0.082$, $\nu =0$ and $H=1/2$.
The dashed line is a fit to the linear behavior
at small areas.
\label{fig:linear}
}
\end{figure}

As noted in the introduction, analytic theories of contact
between self-affine surfaces predict that the real area
of contact $A$ should be proportional to the applied load $W$
at small loads \cite{greenwood66,volmer97,persson01,bush75}.
To make the load dimensionless we normalize
it by $A_0$ times the effective modulus $E'$.
Figure \ref{fig:linear} shows a plot
of the fraction of the projected area that is
in contact $A/A_0$ as a function of the normalized load
for a system with $L=256$, $\Delta =0.082$, $\nu =0$
and $H=1/2$.
As expected, the area is proportional to the load
at small loads.
A growing deviation from this proportionality is evident
as $A/A_0$ increases above 4 or 5 \%.

\begin{figure}
\includegraphics[height=2.9in]{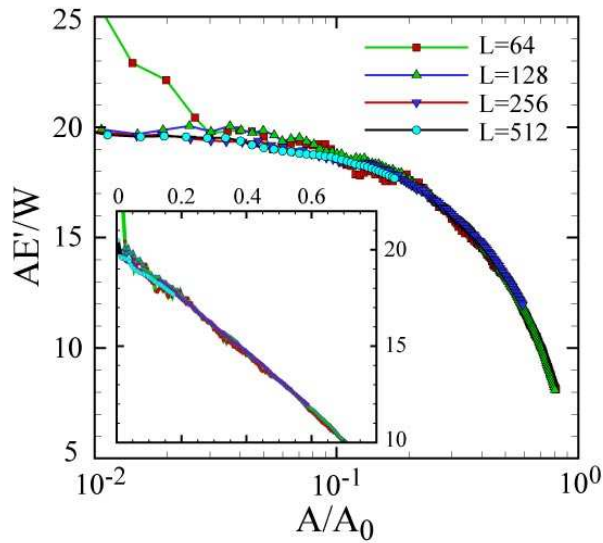}
\caption{
The dimensionless ratio of the area to load
$AE'/W$ vs $\log_{10} (A/A_0)$ for the indicated system sizes.
In all cases $\Delta=0.082$, $\nu=0$ and $H=1/2$.
The inset shows a linear plot of the same data.
\label{fig:ratio}
}
\end{figure}

To emphasize deviations from linearity, the dimensionless
ratio of true contact area to load $AE'/W$
is plotted against $\log_{10} A/A_0 $ in Fig. \ref{fig:ratio}.
Results for $L$ between 64 and 512 fall onto almost
identical curves.
The small variation between curves is comparable to that
between different random surfaces of the same size.
In each case $AE'/W$ is nearly constant
when from 1 to 8\% of the surface is in contact.
This implies that the mean pressure in contacting
regions $\langle p \rangle \equiv W/A$ is also constant.
The ratio $A E'/W$ drops
as $A/A_0$ increases further because
$A/A_0$ is bounded
by unity while the normalized load keeps increasing.
The ratio has roughly halved by $A/A_0 = 70$\%.

Most of the analytic theories mentioned above explicitly
assume that there is a statistically significant number of asperities in
contact, and that only the tops of asperities are in 
contact.
The latter assumption breaks down as $A/A_0$ approaches unity,
contributing to the decrease at large $A/A_0$ in Fig. \ref{fig:ratio}.
The first assumption must break down for our systems
when the total number of nodes in contact,
$L^2 A/A_0$, is small.
This explains the rise in the $L=64$ data for $A/A_0 < 2$\%
in Fig. \ref{fig:ratio}.
This rise is dependent on the specific random
surface generated, and is particularly dramatic for the case
shown.
Examination of this and other data indicates that proportionality
between load and area is observed when there are
more than 100 contacting nodes ($L^2 A/A_0 > 100$ ).

Batrouni {\it et al.}\cite{batrouni02} considered the same
range of system sizes for $\nu=0$.
They fitted all data from $A/A_0 \leq 0.2 $ to a power law and found
$W \propto A^\gamma$ with $\gamma >1$.
Their numerical results clearly rule out
an earlier prediction\cite{roux93}
that $\gamma = (1+H)/2$, but we do not believe that
their results are inconsistent with an initially linear relation between
load and area ($\gamma=1$).
Their fit included regions where the number of contacting
nodes is below the minimum threshold just described.
In addition, their value of $\gamma$ decreased steadily towards
unity with increasing $L$, varying from 1.18 at $L=32$
to 1.08 at $L=256$.
We believe that this small difference from unity is within
the systematic errors associated with the limited scaling range.
Note that $\gamma=1.1$ would imply a 25\% decrease in
Fig. \ref{fig:ratio} from $A/A_0=10^{-2}$ to $10^{-1}$ which
is much larger than the observed change ($\sim 5$\%).
The inset of Fig. \ref{fig:ratio} shows
that data up to $A/A_0=0.7$ can be described by a very
simple linear dependence of $A/W$ on $A$.
We conclude that $\gamma=1$, but that leading quadratic
corrections to scaling give a larger apparent exponent when
a large range of data is fit.

The lack of system-size dependence in Fig. \ref{fig:ratio}
may appear surprising in the context of some previous
results for self-affine surfaces.
These studies \cite{persson01,borri01} considered a fixed roughness
at large scales and examined changes in contact area
with increasing resolution.
They found that the contact area decreased as the number of
nodes increased because the local slope of the surface became
rougher at higher resolution.
Our Fig. \ref{fig:ratio} compares results for
the same small scale roughness and shows that there is
a well-defined thermodynamic limit as one increases the
total system size.
This result is not obvious, since the rms roughness at the
scale of the entire contact rises as $L^H \Delta$.
Apparently this increase in large scale roughness is
irrelevant because it rises sufficiently slowly with $L$.
As noted in Sec. \ref{sec:geometry}, the large scale roughness is sensitive
to the first few random numbers chosen in generating
the self-affine surface.
If surfaces with the same large scale roughness are
compared, substantial differences are found because
the small scale roughness varies.
These fluctuations are absent when results from the
same small scale roughness are compared.

The existence of a well-defined thermodynamic limit
allows us to consider results for a single system
size in subsequent sections, and extrapolate the results
to other cases.
In the following sections we will focus on the constant
region of Fig. \ref{fig:ratio} and
examine variation of this value with the statistical
properties of the interface.
Unless noted, all results are for $L=256$ and
uncertainties due to statistical fluctuations are
less than 5\%.

\subsection{Comparison to analytic theories }

Bush {\it et al.} \cite{bush75} found that the ratio plotted in Fig. \ref{fig:ratio}
should increase inversely with the root mean squared slope
of the surface $\sqrt{< |\nabla h|^2>}$ .
More specifically, they predicted that the quantity
\begin{equation}
\kappa \equiv \sqrt{< |\nabla h|^2>} AE'/W
\label{eq:Bush}
\end{equation}
should have the constant value of $\sqrt{2 \pi}$.
Persson arrived at a rather different looking expression
in terms of the height-height correlation function
(Eq. \ref{eq:cq} and \ref{eq:cr}) \cite{persson01}.
However, it can be reduced to a prediction that
$\kappa = \sqrt{8/ \pi}$ using the fact that
$q^2 C(q)$ is the Fourier transform of $|\nabla h |^2$.
Note that both predictions have a well-defined thermodynamic
limit that is independent of large scale roughness,
just as observed in our simulations.

\begin{figure}
\includegraphics[height=2.9in]{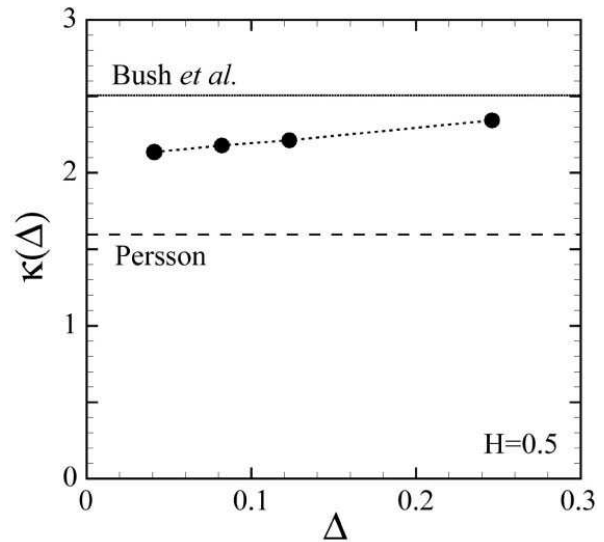}
\caption{
The product $\kappa$ (Eq. \ref{eq:Bush}) as a function of
roughness $\Delta$ for $H=1/2$ and $\nu=0$, and the constant
values predicted by Bush {\it et al.} (solid line) \cite{bush75}
and Persson (dashed line) \cite{persson01}.
The dotted line is a guide to the eye.
\label{fig:delta}
}
\end{figure}

For our surfaces, $\Delta$ is the rms change in height between
adjacent nodes in each of the two spatial directions.
Thus $\sqrt{<|\nabla h|^2>}=\sqrt{2} \Delta$.
Numerical results for $\kappa$ are plotted against
$\Delta$ in Fig. \ref{fig:delta} for $\nu=0$ and $H=1/2$.
The value of $\kappa$ only changes about 10\%,
while the roughness changes by almost an order of magnitude.
If $\Delta$ is increased to larger values than considered here,
the local slope of the surface exceeds unity in some regions.
This regime was not studied because it requires a different
meshing algorithm and
most treatments of contacts assume that the local slope
remains less than unity.

\begin{figure}
\includegraphics[height=2.9in]{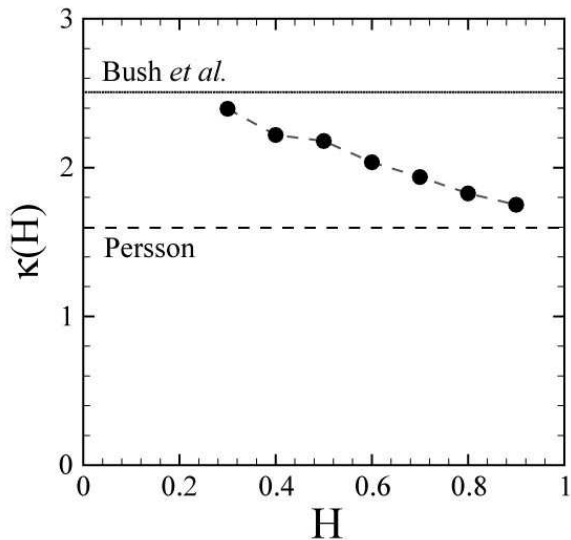}
\caption{
The product $\kappa$ (Eq. \ref{eq:Bush}) as a function
of $H$ for $\Delta=0.082$,
and the constant values predicted by Bush {\it et al.} (solid line)
\cite{bush75}
and Persson (dashed line) \cite{persson01}.
The dotted line is a guide to the eye.
\label{fig:hurst}
}
\end{figure}

We also evaluated $\kappa$ over the range of roughness
exponents typically observed on real surfaces,
$ 0.3 \leq H \leq 0.7$ \cite{krim95,meakin98}, and at the higher
values of 0.8 and 0.9.
Figure \ref{fig:hurst} shows results for
a fixed value of $\Delta = 0.082$.
Even though the large scale roughness, $L^H \Delta$, varies by 
more than an order of magnitude,
$\kappa$ changes by less than 30\%.
There is a nearly linear decrease with increasing $H$
that may be influenced by uncertainties in determining
the true contact area.
As shown in Sec. \ref{sec:area},
increasing $H$ also increases the population
of large clusters and may reduce uncertainties associated
with assuming that
the entire square region around
contacting nodes is in contact.

\begin{figure}
\includegraphics[height=2.8in]{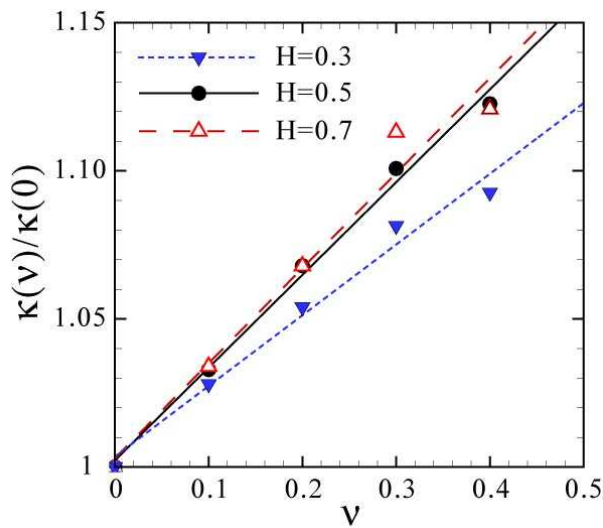}
\caption{The ratio of $\kappa$ to its value at $\nu =0$ as
a function of $\nu$.
Results for $H$ between 0.3 and 0.7 show nearly the same
linear rise with $\nu$.
Lines are linear fits to the data.
}
\label{fig:poisson} 
\end{figure}

Fig. \ref{fig:poisson} shows that the Poisson ratio also
has relatively little effect on $\kappa$.
Results for each value of $H$ are normalized by the
value $\kappa(0)$ obtained at $\nu=0$.
In every case there is a nearly linear rise in $\kappa$
at small $\nu$, that appears to saturate as $\nu$ approaches
the limiting value of 0.5.
The total change of around 10\% is comparable to the
change found with $\Delta$.
The increase in $\kappa$ with $\nu$ appears to be 
related to increased interactions between nearby asperities.
The lateral expansion in response to a normal stress increases
with $\nu$.
This reduces the local curvature, making it easier for adjacent
regions to come into contact.
Detailed analysis of neighboring asperities shows that a
smaller peak between two contacts may be brought up in to contact
at high Poisson ratios.

All of the values of $\kappa$ in Figs. \ref{fig:delta} and \ref{fig:hurst}
lie between the analytic predictions of Bush {\it et al.} and Persson.
Our results suggest that using a value of $\kappa = 2.2$ should
predict the ratio of area to load within about 10\% over a wide
range of surface geometries at $\nu=0$.
Fig. \ref{fig:poisson} indicates that the value of $\kappa$ 
should be increased linearly to about 2.5 as $\nu$ increases
to the limiting value of 0.5.

The agreement with these analytic predictions is quite good
considering the ambiguities in discretization of the surface.
Both analytic models assume that the surface has continuous
derivatives below the small length scale cutoff of the roughness.
Bush {\it et al.} \cite{bush75} consider contact
between elliptical asperities and
Persson \cite{persson01} removes all Fourier content above some wavevector.
While we use quadratic shape functions, the contact algorithm
only considers nodal heights and assumes that
contact of a node implies contact
over the entire corresponding square.
One might expect that this assumption would lead to larger
areas of contact, and our results do lie above those of
Persson.
Discretization would not be important if the spacing between
nodes were much smaller than the typical size of asperity contacts.
However, as we now show, the majority of the contact area consists of
clusters containing only a few nodes.
The number of large clusters grows as $H \rightarrow 1$,
which may explain why our
numerical results approach Persson's prediction in this limit.

\subsection{Distribution of connected contact regions}
\label{sec:area}

Most continuum theories approximate the real contact area by
summing over many disconnected asperity contacts, each of which has
a circular \cite{greenwood66} or elliptical \cite{bush75} shape.
The connected regions in our calculated contacts
(e.g. Fig.  \ref{fig:contact}) are considerably more complicated.
We consider two  nodes to be connected if they
are nearest neighbors on the square lattice of interfacial nodes
\cite{footnext}.
All clusters of connected contacting nodes are then identified
for each load.
The area of each cluster $a_c$ is just the number of connected
nodes, since each represents a square region of unit area.

\begin{figure}
\includegraphics[height=2.8in]{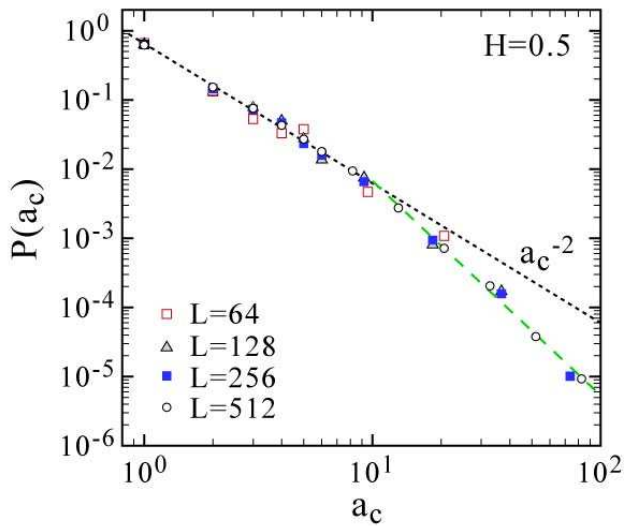}
\caption{Probability $P$ of a connected cluster of area $a_c$
as a function of $a_c$ for $\nu=0$, $H=1/2$, $\Delta=0.082$
and the indicated system sizes.
All results follow a power law, $P(a_c) \sim a_c^{-\tau}$,
with $\tau=3.1$ (dashed) line at large $a_c$.
The dotted line corresponds to $\tau=2$.
}
\label{fig:clustersizeL}
\end{figure}

Figure \ref{fig:clustersizeL} shows the probability of
finding a cluster of a given area $P (a_c)$ as a function
of $a_c$ for $H=0.5$, $\Delta=0.082$ and $\nu=0$.
Results for different system sizes
collapse onto a common curve.
For $a_c > 8$ the curve can be described by a power law
$P (a_c) \sim a_c^{- \tau}$ with $\tau=3.1$
(dashed line).
This rapid fall off ($\tau >1$) means that the integral of $P$ is
dominated by small clusters.
Thus even though the maximum observed cluster size
grows with $L$, the value of $P$ at small $a_c$ is unaffected.
All of the data shown in Fig. \ref{fig:clustersizeL}
are for $A/A_0$ between 5 and 10\%,
but we find that the distribution of clusters is nearly
constant for $A/A_0 \le 10\%$.
This is the same range where $A$ and load are
nearly linearly related.
The probability of large clusters rises markedly for $A/A_0 > 0.3$,
as clusters begin to merge and eventually
percolate across the interface.
The following data is all for
$A/A_0 \le 0.1$.

The model considered by Greenwood and Williamson also gives a load
independent $P(a_c)$ and this is a central reason for the linear
relation between load and area in their model.
As the load increases, each existing cluster grows larger and new small
clusters are generated in a way that maintains a stationary distribution
of cluster sizes.
Only the total number of clusters changes, and it rises linearly with load.
Our calculated $P(a_c)$ and the total number of clusters both
follow this behavior. 
However, the distribution and the shapes of the clusters are very
different than assumed by Greenwood and Williamson.

\begin{figure}
\includegraphics[height=2.8in]{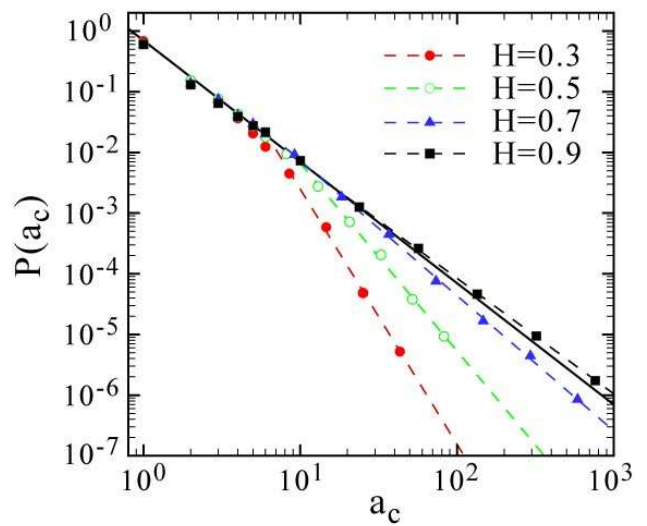}
\caption{Probability $P$ of a connected cluster
as a function of area $a_c$ for $\nu=0$, $\Delta=0.082$, $L=512$
and the indicated values of $H$.
Dashed lines indicate the asymptotic power law behavior
with $\tau=4.2$, 3.1, and 2.3 for $H=0.3$, 0.5 and 0.7,
respectively.
The solid line corresponds to $\tau=2$.
}
\label{fig:clustersizeH}
\end{figure}

The variation of $P(a_c)$ with $H$ is
shown in Fig. \ref{fig:clustersizeH}.
Results for very small clusters ($a_c <8$) are nearly
independent of $H$, but the asymptotic power law behavior
at large $a_c$
changes dramatically.
The range of scaling behavior is too small for precise
determination of $\tau$, but our data is consistent with
$\tau=3.1 \pm 0.2$ for $H=0.5$.
When $H < 0.5$ there is an anticorrelation between the surface
slopes in nearby regions \cite{meakin98}.
This leads to more rapid up and down fluctuations that make
large contacts unlikely and yield a larger $\tau = 4.2 \pm 0.4$
for $H=0.3$.
When $H > 0.5$ there is a positive correlation between local
slopes, yielding larger clusters.
Fits give $\tau=2.3 \pm 0.2$ for $H=0.7$ and
$\tau=1.9\pm 0.1$ for $H=0.9$.
For $H < 0.9$ we find $\tau >2$, which implies that the
mean cluster size $\langle a_c \rangle $ is dominated by small clusters.
Directly calculated sizes are indeed independent of both $L$ and $A/A_0$.
We find $\langle a_c \rangle = 1.8$, 2.5, and 4.0
for $H=0.3$, 0.5 and 0.7, respectively.
For $H=0.9$, $\tau$ appears to be slightly smaller than 2,
and the mean cluster size grows weakly with $L$.

Some approximate treatments of contact begin by assuming that
the two surfaces do not deform and then determine the
regions where the two solids would
interpenetrate \cite{dieterich94,dieterich96,greenwood01}.
Scaling arguments \cite{meakin98} and simulations \cite{meakin98,kondev95}
show that this rigid overlap model
gives a power law distribution of connected areas at large $a_c$
with $\tau =2-H/2$. 
This is qualitatively consistent with Dieterich and Kilgore's
experiments where $\tau$ varied from 1 to 2 and tended
to decrease with increasing $H$.
However, it is qualitatively different from our results where
$\tau$ is always greater than 2.
One consequence is that the mean cluster size from the
rigid overlap model diverges with system size as $L^H$,
while it remains of order the discretization size in
our calculation \cite{footnext}.

\begin{figure}
\includegraphics[height=2.9in]{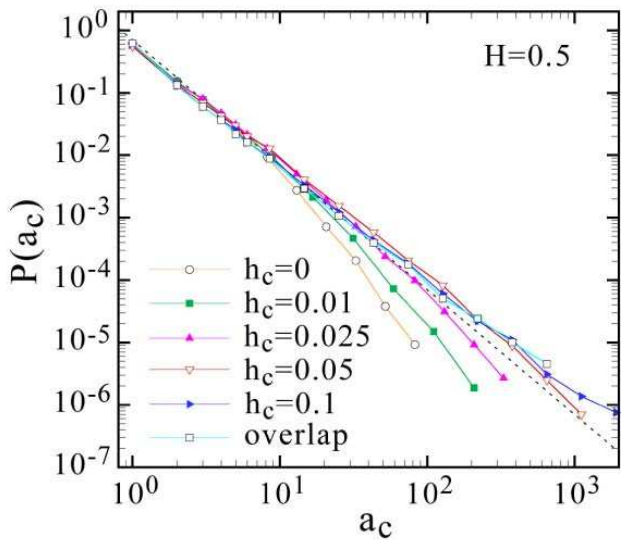}
\caption{
Probability $P$ of a connected cluster as a function of area $a_c$
for $\nu=0$, $H=1/2$, $\Delta=0.082$, $L=512$ and different criteria
for contact.
The probability distribution
for the rigid overlap model (open squares)
falls off more slowly than $a_c^{-2}$ (dotted line).
When only contacting nodes are included ($h_c=0$), $P$ falls
off more rapidly (open circles).
As the width $h_c$ of the region considered in contact
increases, results from the full calculation approach the
overlap results.
}
\label{fig:clustersizerig}
\end{figure}

A possible explanation for the discrepancy between our results
and experiment is that the latter identifies regions that
are within some small fraction of the wavelength of light
as being in contact.
Figure \ref{fig:clustersizerig} illustrates the dramatic
effect that this can have on $P(a_c)$. 
The uppermost curve shows the cluster distribution obtained
by applying the rigid overlap model to our surfaces.
The asymptotic slope is consistent with 
the analytic prediction for $H=0.5$: $\tau=2-H/2=1.75$.
The lowermost curve is our result for the actual
contact area.
The intermediate curves were obtained by changing our definition
of contact to include all nodes that are separated by less
than some value $h_c$.
As $h_c$ increases, the number of nodes in the contact
region rises and the probability of large clusters grows.
When $h_c$ is comparable to or larger than $\Delta=0.082$,
$P(a_c)$ follows the rigid overlap prediction quite closely.
It is likely that the optical experiments were in this
limit.
However, it is also likely that plastic deformation is
important in these experiments.
This effect will be explored in future work.

\subsection{Contact Morphology}

\begin{figure}
\includegraphics[height=5.20in]{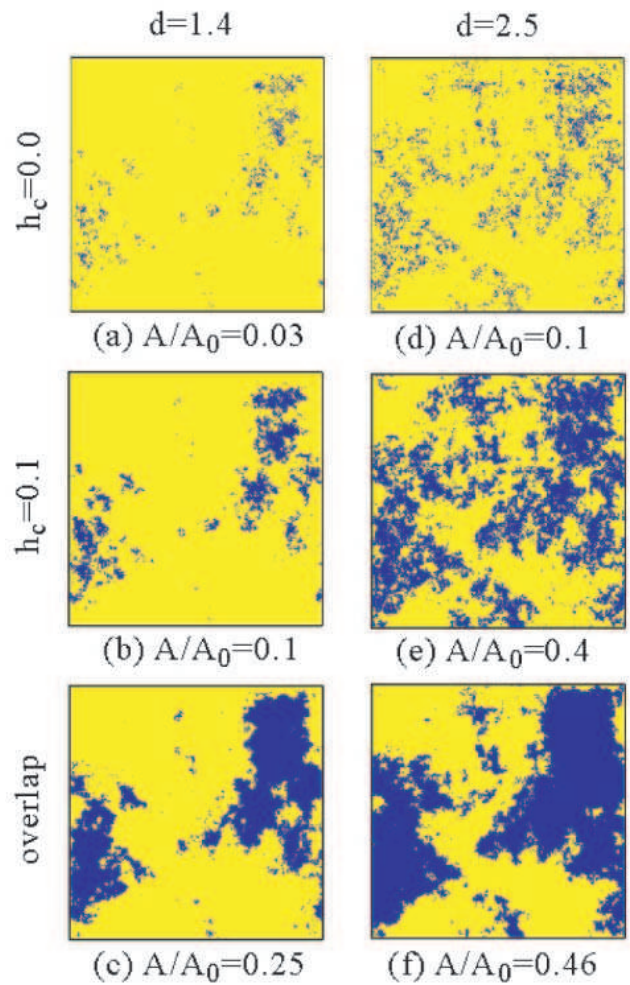}
\caption{Contact morphology for $L=256$ under two different
values of the displacement $d$ following contact: $d=1.4, 2.54$.
Panels (a,d) show regions in true contact ($h_c=0$),
while (b,e) show regions where the surface separation is
less than $h_c=0.1$.
Panels (c,f) show the contacts predicted by the rigid overlap
model.
The fractional contact area ($A/A_0$) is indicated for each case.
}
\label{fig:contactpics}
\end{figure}

The contact morphologies produced by different models are contrasted
in Fig. \ref{fig:contactpics}.
Results for two values of the interpenetration $d$ are shown,
where $d$ is the downward displacement applied to the top
of the elastic solid (Fig. \ref{fig:FEM})
after the surfaces first touch.
The top panels show the results of the full calculation,
the middle panels are for $h_c=0.1$, and
the bottom panels show the contacts obtained from the
rigid overlap model.
The first obvious difference between the results is that the
rigid overlap grossly overestimates the fraction of area in
contact.
For small $d$ the actual area is roughly 8 times smaller than
given by the overlap model.

\begin{figure}
\includegraphics[height=3.0in]{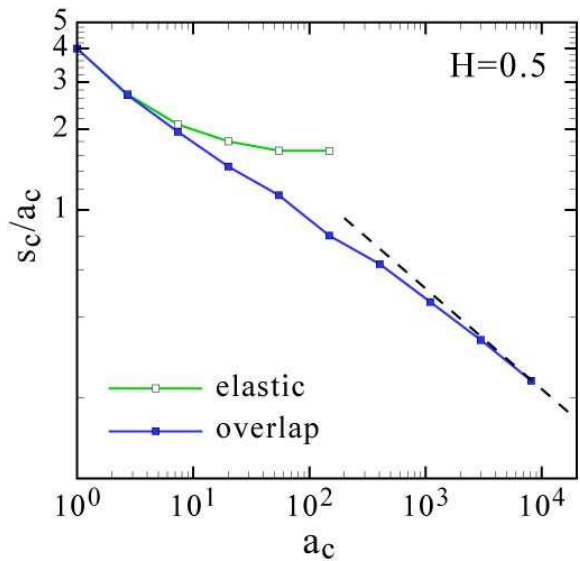}
\caption{
Ratio of perimeter $s_c$ to area $a_c$ vs. area for
the full calculation (open squares) and rigid overlap model (filled squares).
The dashed line shows the asymptotic prediction for the
overlap model.
}
\label{fig:sovera}
\end{figure}

As seen above, the
rigid overlap model also gives many more large clusters.
Moreover the shapes of the clusters are quite different.
Analytic studies predict that the overlap model should give
clusters with a non-fractal
interior, but with a fractal interface.
More specifically, if $R$ is the diameter of a cluster,
then the area $a_c \propto R^2$, but the perimeter length
$s_c \propto R^{D_f}$
where the fractal dimension $D_f=(3-H)/2$ \cite{meakin98,kondev95}.
Thus as the clusters grow in size, the perimeter becomes smaller
and smaller relative to the area: $s_c/a_c \propto a_c^{-(1+H)/4}$.
Figure \ref{fig:sovera} shows that our results for the
overlap model are consistent with this scaling prediction.
However, the results for the full calculation are quite different.
The value of $s_c/a_c$ approaches a constant at large $a_c$
indicating that the perimeter and area are both fractals
with the same fractal dimension.
Plots of $s_c$ and $a_c$ as a function of $R$ show $D_f$ is
roughly 1.6 for $H=0.5$.
Note that $s_c$ is actually larger than $a_c$ because it is
defined as the number of missing nearest-neighbors
along the periphery.
Thus it would be 4 for a cluster containing a single node.

Results for other values of $H$ look very similar to those
for $H=0.5$.
In each case $s_c/a_c$ saturates at large $a_c$, indicating
the area and perimeter have the same fractal dimension.
The limiting value of $s_c/a_c$ decreases from about 1.7 for
$H=0.3$ to 1.1 for $H=0.9$.
The large number of perimeter nodes leads to some ambiguity
in the total area obtained from our calculation.
If only a fraction of the square around each node were
actually in contact, then the true contact area would
be smaller, moving our results in Figs. \ref{fig:delta}
and \ref{fig:hurst} closer to Persson's result
\cite{persson01,persson02}.
It is interesting that our results approach Persson's prediction
as $H \rightarrow 1$, and the perimeter becomes less important.

Despite the above differences, the rigid overlap model does
provide information about where real contacts may occur.
The distance between surfaces is always larger than that given by
the overlap model because $\nu=0$.
Thus all of the contacts in panels (a) and (d) are part of the
overlapping regions shown in (c) and (f).
Only a fraction of the overlapping regions is in contact, because
a local peak can screen a neighboring valley from contact.
As pointed out by Greenwood \cite{greenwood84}, a large fraction of
points are local maxima, so the average cluster size is
comparable to the lattice resolution.
On the other hand, a small local maximum can only screen
a small local region.
Thus there tend to be many small contacts in the regions where
overlap first occurs.
These points lie in the middle of the large clusters in
panels (c) and (f).
A higher density of clusters, and clusters of larger size,
are found in these regions of panels (a) and (d).
As noted above, when nodes that are within a distance comparable
to $\Delta$ are considered in contact, the distribution of
clusters approaches that for the rigid overlap model.
Panels (b) and (e) show the contact morphology produced in
this limit ($h_c=0.1$).
Note that the clusters from panels (a) and (d) have been
connected into larger clusters that still lie within those
of panels (c) and (f).
Growth is most pronounced in regions where overlap is greatest.
These regions carry a greater share of the load and flatten
more.

Greenwood and Wu \cite{greenwood01} have recently reconsidered
when a local peak should be considered as an asperity.
They conclude that one should think of each cluster in the rigid
overlap model as a single asperity, and use the diameter and
height of the overlap to determine the dimensions of an effective
ellipsoidal asperity.
Our results indicate that the original view \cite{greenwood84}
that almost all points
are asperities and the typical
asperity diameter is comparable to
the lattice size provides a more accurate description of the contact.
However, the revised approach of identifying overlapping regions
with asperities may give a better description of
subsurface stresses, because it captures correlations in the location
of load bearing regions.
Since the maximum shear stress is usually below the surface,
the overlap model \cite{greenwood01} may be useful in modeling wear.

\subsection{Distribution of local pressures}

\begin{figure}
\includegraphics[height=2.9in]{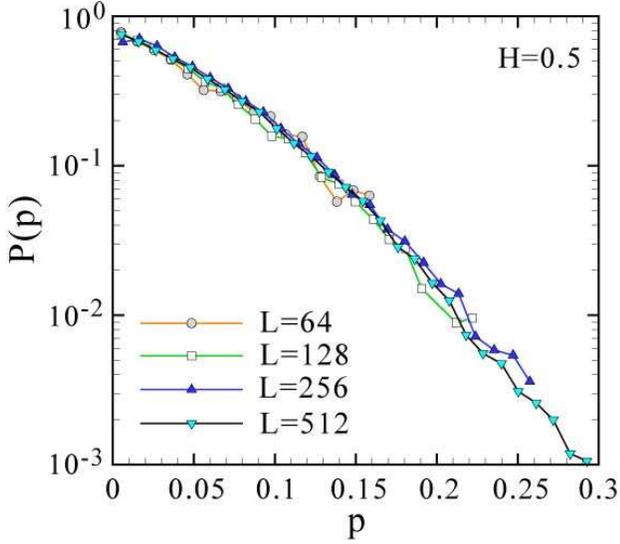}
\caption{
Probability distribution of the local pressure at contacting
nodes for different system sizes with
$\Delta=0.082$, $\nu=0$, $H=0.5$,
and $A/A_0$ between 5 and 10 \%.
}
\label{fig:pressure}
\end{figure}

Plastic deformation at the interface will be influenced by
how pressure is distributed within the contact area.
We find that this distribution has a strikingly universal form.
Figure \ref{fig:pressure} shows that the probability $P(p)$ for
a contacting node to have local pressure $p$
is independent of system size.
Since the contact area increases linearly with load,
the mean local contact pressure $\langle p \rangle = W/A$
is independent of contact area, and 
the entire distribution also remains unchanged
for $A/A_0$ between about .01 and .1.

\begin{figure}
\includegraphics[height=2.8in]{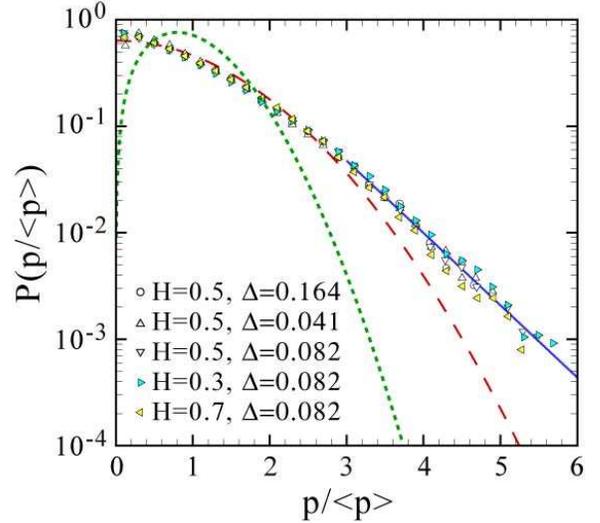}
\caption{
Probability distributions for $p/\langle p \rangle$
at the indicated values of $\Delta$ and $H$ all collapse onto
a universal curve.
Here $\nu=0$ and $A/A_0$ is between 5 and 10\%.
The solid line is a fit to the exponential tail of the distribution,
the dotted line shows Eq. (14),
and the dashed line shows a Gaussian with the appropriate
normalization and mean.
}
\label{fig:universal}
\end{figure}

Increasing the small scale roughness ($\Delta$)
leads to a proportional increase in $\langle p \rangle$.
Yet
Figure \ref{fig:universal} shows that results for all $\Delta$ and $H$
collapse onto a universal function of the
dimensionless variable $p/\langle p \rangle$.
The probability decreases monotonically with increasing
$p$, and for $p/ \langle p \rangle > 3$ follows an exponential decay
(solid line), 
$P(p) \propto \exp (-p/p_1)$,
with $p_1 \approx \langle p \rangle/1.6$.
This exponential tail implies that some regions have stresses
much higher than $\langle p \rangle$ and may undergo
plastic deformation even when the mean stress is much less
than the hardness.

Similar universal curves have been found for the stress distribution
in a variety of ``jammed'' systems \cite{liu01},
including granular media \cite{liu95,coppersmith96},
thermal glasses \cite{ohern01}
and polymer crazes \cite{rottler02}.
In each case the tail of the distribution follows a simple exponential
rather than the Gaussian that might be expected from equilibrium
arguments.
Several explanations for the exponential form have been proposed
\cite{liu95,coppersmith96,ohern01,rottler02},
but most do not apply to our zero temperature, deterministic and
elastic system.
However, it is possible that the power law correlations in interface
height may lead to a hierarchical distribution of load that
is analogous to the $q$-model \cite{coppersmith96}.

The distribution of local pressures plays a 
central role in Persson's theory of contact between self-affine
surfaces \cite{persson01,persson02}.
He defines a resolution $\zeta$ corresponding to the number
of points along an axis at which the height of the surface
is known, and assumes a smooth interpolation between these
points.
Increasing $\zeta$ corresponds to resolving more of the
surface roughness and increases $\Delta$ if the surface
is self-affine.
The pressure distribution $P(p,\zeta)$
is a function of both resolution
and pressure.
Its derivatives satisfy
\begin{equation}
\frac{ \partial P} { \partial \zeta} = G'(\zeta) p_0^2
\frac{\partial^2 P} {\partial p^2}
\label{eq:Persson}
\end{equation}
where $p_0=W/A_0$ is the apparent mean pressure,
primes denote a derivative and
\begin{equation}
G(\zeta)=(E'/p_0)^2 \langle | \bf{\nabla} h |^2 \rangle   .
\end{equation}
Persson obtained solutions for $P(p,\zeta)$ in the geometry
considered here by starting from perfectly flat planes
with $P(p,1)=\delta(p-p_0)$ and iterating to higher
resolution.
He also imposed the boundary condition that the
probability goes to zero at zero pressure \cite{persson02}.

Given the results shown in Fig. 15, it is interesting to ask
if Persson's equations have a universal solution in the limit of
small loads.
Fig. 15 shows the probability distribution within the contact,
while Persson's $P(p,\zeta)$ includes noncontacting regions and its
integral over pressure is the fractional contact area.
Thus it should be related to the universal distribution $\tilde{P}$ by:
$P(p,\zeta)= \tilde{P}(p/<p>) p_0/<p>^2$, where $p_0/<p>=A/A_0$.
Then use of Eqs. (10) and (11) leads to an equation for $\tilde{P}$:
\begin{equation}
2 \tilde P(x) + x \tilde P'(x) + \kappa^2 \tilde P "(x)/4=0  .
\end{equation}
There is a solution with unit norm and mean for Persson's value
of $\kappa = \sqrt{8/\pi}$:
\begin{equation}
\tilde{P}(x)= \frac{\pi}{2} x \exp (-\frac{\pi}{4} x^2)  .
\end{equation}
As shown in Fig. 15, this solution (dotted line) is much more
strongly peaked than our numerical results,
decaying to zero linearly in the limit $p \rightarrow 0$ and
as a Gaussian at large $p$.
Note that since the mean and norm are the same for all curves,
the presence of extra weight at large $p$ implies more weight
at low $p$.
Also shown in Fig. 15 is a pure Gaussian with unit norm and mean (dashed line).
This solution provides a better fit to the numerical data at low $p$.
However,
as for ``jammed'' systems \cite{liu95,coppersmith96,liu01,ohern01,rottler02},
the tail of the numerical distribution is much closer to a pure
exponential than a Gaussian decay.
This may reflect correlations in the loads carried by
different asperities that are not fully captured in analytic theories
\cite{bush75,persson01,persson02,greenwood66}

\section{Summary and Conclusions}
\label{sec: conclusion}

In this paper, we developed a numerical framework for
analyzing frictionless, non-adhesive contacts between self-affine surfaces using
the finite element method.
This method has been applied to perfectly elastic contacts
with a range of Poisson ratios, roughness amplitudes
and roughness exponents.
In each case the real contact area $A$ rises linearly with load $W$
until the fraction of the total area in contact reaches 5 to 10\%
(Fig. \ref{fig:ratio}).
This implies that the average local pressure in the contacts,
$\langle p \rangle = W/A$,
remains constant.
The dimensionless pressure $\langle p \rangle /E'$ 
is independent of the system size even though the large scale
roughness grows as $L^H$.

As predicted by analytic studies \cite{bush75,persson01},
the dimensionless pressure scales linearly with the small
scale roughness $\Delta$.
The constant factor $\kappa$ (Eq. \ref{eq:Bush})
that relates the roughness to the local
pressure is always
between the predictions of Bush {\it et al.} \cite{bush75} and
Persson \cite{persson01}
(Figs. \ref{fig:delta}--\ref{fig:poisson}).
A value of 
$\kappa=2.2$ reproduces the numerical results within about 10\% for $\nu=0$,
and the best fit value rises linearly to about 2.5 as $\nu$ rises
to the limiting value of 0.5.
These results allow the mean pressure and fractional contact
area to be predicted for any elastic self-affine surface with known
small scale roughness.

The detailed morphology of the contact region, and distribution
of the areas $a_c$ of connected regions were also studied.
As in early theories of contact \cite{greenwood66,bush75},
the increase in area with load reflects a linear increase
in the total number of contacts with no change in the
probability distribution of contact areas.
As $W$ increases, each existing contact grows, and new
contacts are formed at a rate that maintains a constant $P(a_c)$.
At large $a_c$, the probability distribution falls off
as a power law: $P(a_c)\propto a_c^{-\tau}$
(Fig. \ref{fig:clustersizeH}).
Since $\tau > 2$ for $H<0.9$,
the mean cluster area is independent of $L$ and comparable
to the resolution of the calculation.

The above results for connected clusters
are consistent with the conclusion that
a large fraction of nodes on a self-affine surface are
local maxima that should be treated as asperities \cite{greenwood84}.
However,
recent experimental \cite{dieterich94,dieterich96}
and theoretical \cite{greenwood01} papers have suggested
a different view.
They examine regions where undeformed surfaces would
overlap and associate each with a contact.
This model gives qualitatively different distributions of areas
(Fig. \ref{fig:clustersizerig}).
The value of $\tau$ is always less than 2, and the mean
cluster area diverges as a power of system size.
The geometry of the clusters is also very different.
The rigid overlap model gives two-dimensional clusters
with fractal perimeters,
while the full calculation gives fractal cluster
areas with the same fractal dimension as the perimeter
(Fig. \ref{fig:sovera}).
Including regions where the surfaces are separated by less
than $h_c$ as part of the contact leads
to dramatic changes in the cluster distribution exponent $\tau$ and
total area.
The results approach the overlap distribution when $h_c$
is comparable to the small scale roughness.
Optical experiments will detect gaps that are much
smaller than the wavelength as in contact and this
may explain why small values of $\tau$ are observed.

Plastic deformation will occur when the local pressure in a contact
exceeds the hardness of the material.
The linear relation between mean pressure and small scale roughness
(Eq. \ref{eq:Bush}) can be used to estimate when this will happen.
The largest experimental values of $p/E'$ are of order $0.1$
and are obtained in amorphous and nanocrystalline materials.
Thus Equation \ref{eq:Bush} implies that deformation
can only be elastic when
$\sqrt{\langle|\nabla h | ^2\rangle}< 0.1 \kappa \sim 0.2$.
This condition is violated for many surfaces, and the
much smaller
hardness of macroscopic crystals
will lead to even tighter constraints on the roughness.
Our approach is readily extended to include plastic deformation,
which will be the subject of future work.

Plastic deformation may occur well before the mean pressure
reaches the hardness because some nodes have local pressures
much larger than $\langle p \rangle$.
Results for all parameters collapse on to a universal probability
distribution $\tilde {P} (p/ \langle p \rangle )$ (Fig. \ref{fig:universal}).
Persson has presented approximate analytic equations for the pressure
distribution.
This analytic distribution drops as a Gaussian at large $p$,
while the numerical results have an exponential tail
that greatly increases the number of sites with large pressures.
Similar exponential distributions are found
in many jammed systems such as sandpiles, glasses and crazes
\cite{liu95,coppersmith96,liu01,ohern01,rottler02}.
A common feature of these systems is a highly nonuniform distribution
of stress.
It is possible that the presence of small bumps on bigger bumps
on still bigger bumps in our systems leads to stress transmission like
that in the $q$-model for sandpiles \cite{liu95,coppersmith96}.
This hierarchical structure may produce stress correlations
that are not included in the analytic model \cite{persson01}.

There are many interesting avenues for future research.
The approach outlined here is readily extensible to include more complex
surface morphologies, plasticity, interfacial friction
and tangential loading of the solids.
More challenging issues include adhesion and the role of atomic
scale roughness.
These issues will require hybrid algorithms that include 
atomic information about interfacial interactions.

\begin{acknowledgments}

This material is based on work supported by the National Science
Foundation under Grant No. CTS-0103408.
We thank N. Bernstein, M. Ciavarella, J. A. Harrison, J. Kondev,
and M. Ciavarella for useful discussions.

\end{acknowledgments}



\begin{thebibliography}{29}
\expandafter\ifx\csname natexlab\endcsname\relax\def\natexlab#1{#1}\fi
\expandafter\ifx\csname bibnamefont\endcsname\relax
  \def\bibnamefont#1{#1}\fi
\expandafter\ifx\csname bibfnamefont\endcsname\relax
  \def\bibfnamefont#1{#1}\fi
\expandafter\ifx\csname citenamefont\endcsname\relax
  \def\citenamefont#1{#1}\fi
\expandafter\ifx\csname url\endcsname\relax
  \def\url#1{\texttt{#1}}\fi
\expandafter\ifx\csname urlprefix\endcsname\relax\def\urlprefix{URL }\fi
\providecommand{\bibinfo}[2]{#2}
\providecommand{\eprint}[2][]{\url{#2}}

\bibitem[{\citenamefont{Bowden and Tabor}(1986)}]{bowden86}
\bibinfo{author}{\bibfnamefont{F.~P.} \bibnamefont{Bowden}} \bibnamefont{and}
  \bibinfo{author}{\bibfnamefont{D.}~\bibnamefont{Tabor}},
  \emph{\bibinfo{title}{The Friction and Lubrication of Solids}}
  (\bibinfo{publisher}{Clarendon Press}, \bibinfo{address}{Oxford},
  \bibinfo{year}{1986}).

\bibitem[{\citenamefont{Dieterich and Kilgore}(1994)}]{dieterich94}
\bibinfo{author}{\bibfnamefont{J.~H.} \bibnamefont{Dieterich}}
  \bibnamefont{and} \bibinfo{author}{\bibfnamefont{B.~D.}
  \bibnamefont{Kilgore}}, \bibinfo{journal}{Pure and Applied Geophysics}
  \textbf{\bibinfo{volume}{143}}, \bibinfo{pages}{283} (\bibinfo{year}{1994}).

\bibitem[{\citenamefont{Dieterich and Kilgore}(1996)}]{dieterich96}
\bibinfo{author}{\bibfnamefont{J.~H.} \bibnamefont{Dieterich}}
  \bibnamefont{and} \bibinfo{author}{\bibfnamefont{B.~D.}
  \bibnamefont{Kilgore}}, \bibinfo{journal}{Tectonophysics}
  \textbf{\bibinfo{volume}{256}}, \bibinfo{pages}{219} (\bibinfo{year}{1996}).

\bibitem[{\citenamefont{Berthoud and Baumberger}(1998)}]{berthoud98}
\bibinfo{author}{\bibfnamefont{P.}~\bibnamefont{Berthoud}} \bibnamefont{and}
  \bibinfo{author}{\bibfnamefont{T.}~\bibnamefont{Baumberger}},
  \bibinfo{journal}{Proc. R. Soc. Lond A} \textbf{\bibinfo{volume}{454}},
  \bibinfo{pages}{1615} (\bibinfo{year}{1998}).

\bibitem[{\citenamefont{Greenwood}(1984)}]{greenwood84}
\bibinfo{author}{\bibfnamefont{J.~A.} \bibnamefont{Greenwood}},
  \bibinfo{journal}{Proc. R. Soc. Lond. A} \textbf{\bibinfo{volume}{393}},
  \bibinfo{pages}{133} (\bibinfo{year}{1984}).

\bibitem[{\citenamefont{Greenwood and Wu}(2001)}]{greenwood01}
\bibinfo{author}{\bibfnamefont{J.~A.} \bibnamefont{Greenwood}}
  \bibnamefont{and} \bibinfo{author}{\bibfnamefont{J.~J.} \bibnamefont{Wu}},
  \bibinfo{journal}{Meccanica} \textbf{\bibinfo{volume}{36}},
  \bibinfo{pages}{617} (\bibinfo{year}{2001}).

\bibitem[{\citenamefont{Volmer and Natterman}(1997)}]{volmer97}
\bibinfo{author}{\bibfnamefont{A.}~\bibnamefont{Volmer}} \bibnamefont{and}
  \bibinfo{author}{\bibfnamefont{T.}~\bibnamefont{Natterman}},
  \bibinfo{journal}{Z. Phys. B} \textbf{\bibinfo{volume}{104}},
  \bibinfo{pages}{363} (\bibinfo{year}{1997}).

\bibitem[{\citenamefont{Persson}(2001)}]{persson01}
\bibinfo{author}{\bibfnamefont{B.~N.~J.} \bibnamefont{Persson}},
  \bibinfo{journal}{Phys. Rev. Lett.} \textbf{\bibinfo{volume}{87}},
  \bibinfo{pages}{116101} (\bibinfo{year}{2001}).

\bibitem[{\citenamefont{Persson et~al.}(2002)\citenamefont{Persson, Bucher, and
  Chiaia}}]{persson02}
\bibinfo{author}{\bibfnamefont{B.~N.~J.} \bibnamefont{Persson}},
  \bibinfo{author}{\bibfnamefont{F.}~\bibnamefont{Bucher}}, \bibnamefont{and}
  \bibinfo{author}{\bibfnamefont{B.}~\bibnamefont{Chiaia}},
  \bibinfo{journal}{Phys. Rev. B} \textbf{\bibinfo{volume}{65}},
  \bibinfo{pages}{184106} (\bibinfo{year}{2002}).

\bibitem[{\citenamefont{Batrouni et~al.}(2002)\citenamefont{Batrouni, Hansen,
  and Schmittbuhl}}]{batrouni02}
\bibinfo{author}{\bibfnamefont{G.~G.} \bibnamefont{Batrouni}},
  \bibinfo{author}{\bibfnamefont{A.}~\bibnamefont{Hansen}}, \bibnamefont{and}
  \bibinfo{author}{\bibfnamefont{J.}~\bibnamefont{Schmittbuhl}},
  \bibinfo{journal}{Europhys. Lett.} \textbf{\bibinfo{volume}{60}},
  \bibinfo{pages}{724} (\bibinfo{year}{2002}).

\bibitem[{\citenamefont{Ciavarella et~al.}(2000)\citenamefont{Ciavarella,
  Demelio, Barber, and Jang}}]{ciavarella00}
\bibinfo{author}{\bibfnamefont{M.}~\bibnamefont{Ciavarella}},
  \bibinfo{author}{\bibfnamefont{G.}~\bibnamefont{Demelio}},
  \bibinfo{author}{\bibfnamefont{J.~R.} \bibnamefont{Barber}},
  \bibnamefont{and} \bibinfo{author}{\bibfnamefont{Y.~H.} \bibnamefont{Jang}},
  \bibinfo{journal}{Proc. R. Soc. London A} \textbf{\bibinfo{volume}{456}},
  \bibinfo{pages}{387} (\bibinfo{year}{2000}).

\bibitem[{\citenamefont{Roux et~al.}(1993)\citenamefont{Roux, Schmittbuhl,
  Vilotte, and Hansen}}]{roux93}
\bibinfo{author}{\bibfnamefont{S.}~\bibnamefont{Roux}},
  \bibinfo{author}{\bibfnamefont{J.}~\bibnamefont{Schmittbuhl}},
  \bibinfo{author}{\bibfnamefont{J.-P.} \bibnamefont{Vilotte}},
  \bibnamefont{and} \bibinfo{author}{\bibfnamefont{A.}~\bibnamefont{Hansen}},
  \bibinfo{journal}{Europhys. Lett.} \textbf{\bibinfo{volume}{23}},
  \bibinfo{pages}{277} (\bibinfo{year}{1993}).

\bibitem[{\citenamefont{Johnson}(1985)}]{johnson85}
\bibinfo{author}{\bibfnamefont{K.~L.} \bibnamefont{Johnson}},
  \emph{\bibinfo{title}{Contact Mechanics}} (\bibinfo{publisher}{Cambridge},
  \bibinfo{address}{New York}, \bibinfo{year}{1985}).

\bibitem[{\citenamefont{Greenwood and Williamson}(1966)}]{greenwood66}
\bibinfo{author}{\bibfnamefont{J.~A.} \bibnamefont{Greenwood}}
  \bibnamefont{and} \bibinfo{author}{\bibfnamefont{J.~B.~P.}
  \bibnamefont{Williamson}}, \bibinfo{journal}{Proc. Roy. Soc. A}
  \textbf{\bibinfo{volume}{295}}, \bibinfo{pages}{300} (\bibinfo{year}{1966}).

\bibitem[{\citenamefont{Bush et~al.}(1975)\citenamefont{Bush, Gibson, and
  Thomas}}]{bush75}
\bibinfo{author}{\bibfnamefont{A.~W.} \bibnamefont{Bush}},
  \bibinfo{author}{\bibfnamefont{R.~D.} \bibnamefont{Gibson}},
  \bibnamefont{and} \bibinfo{author}{\bibfnamefont{T.~R.}
  \bibnamefont{Thomas}}, \bibinfo{journal}{Wear} \textbf{\bibinfo{volume}{35}},
  \bibinfo{pages}{87} (\bibinfo{year}{1975}).

\bibitem[{\citenamefont{Mandelbrot}(1979)}]{mandelbrot}
\bibinfo{author}{\bibfnamefont{B.~B.} \bibnamefont{Mandelbrot}},
  \emph{\bibinfo{title}{The fractal geometry of nature}}
  (\bibinfo{publisher}{W. H. Freeman}, \bibinfo{address}{New York},
  \bibinfo{year}{1979}).

\bibitem[{\citenamefont{Krim and Palasantzas}(1995)}]{krim95}
\bibinfo{author}{\bibfnamefont{J.}~\bibnamefont{Krim}} \bibnamefont{and}
  \bibinfo{author}{\bibfnamefont{G.}~\bibnamefont{Palasantzas}},
  \bibinfo{journal}{Int. J. of Modern Phys. B} \textbf{\bibinfo{volume}{9}},
  \bibinfo{pages}{599} (\bibinfo{year}{1995}).

\bibitem[{\citenamefont{Meakin}(1998)}]{meakin98}
\bibinfo{author}{\bibfnamefont{P.}~\bibnamefont{Meakin}},
  \emph{\bibinfo{title}{Fractals, scaling and growth far from equilibrium}}
  (\bibinfo{publisher}{Cambridge}, \bibinfo{address}{Cambridge},
  \bibinfo{year}{1998}).

\bibitem[{\citenamefont{Borri-Brunetto
  et~al.}(2001)\citenamefont{Borri-Brunetto, Chiaia, and Ciavarella}}]{borri01}
\bibinfo{author}{\bibfnamefont{M.}~\bibnamefont{Borri-Brunetto}},
  \bibinfo{author}{\bibfnamefont{B.}~\bibnamefont{Chiaia}}, \bibnamefont{and}
  \bibinfo{author}{\bibfnamefont{M.}~\bibnamefont{Ciavarella}},
  \bibinfo{journal}{Computer Methods in Applied Mechanics and Engineering}
  \textbf{\bibinfo{volume}{190}}, \bibinfo{pages}{6053} (\bibinfo{year}{2001}).

\bibitem[{\citenamefont{h.~Liu et~al.}(1995)\citenamefont{h.~Liu, Nagel,
  Schecter, Coppersmith, Majumdar, Narayan, and Witten}}]{liu95}
\bibinfo{author}{\bibfnamefont{C.}~\bibnamefont{h.~Liu}},
  \bibinfo{author}{\bibfnamefont{S.~R.} \bibnamefont{Nagel}},
  \bibinfo{author}{\bibfnamefont{D.~A.} \bibnamefont{Schecter}},
  \bibinfo{author}{\bibfnamefont{S.~N.} \bibnamefont{Coppersmith}},
  \bibinfo{author}{\bibfnamefont{S.}~\bibnamefont{Majumdar}},
  \bibinfo{author}{\bibfnamefont{O.}~\bibnamefont{Narayan}}, \bibnamefont{and}
  \bibinfo{author}{\bibfnamefont{T.~A.} \bibnamefont{Witten}},
  \bibinfo{journal}{Science} \textbf{\bibinfo{volume}{269}},
  \bibinfo{pages}{513} (\bibinfo{year}{1995}).

\bibitem[{\citenamefont{Coppersmith et~al.}(1996)\citenamefont{Coppersmith,
  h.~Liu, Majumdar, Narayan, and Witten}}]{coppersmith96}
\bibinfo{author}{\bibfnamefont{S.~N.} \bibnamefont{Coppersmith}},
  \bibinfo{author}{\bibfnamefont{C.}~\bibnamefont{h.~Liu}},
  \bibinfo{author}{\bibfnamefont{S.}~\bibnamefont{Majumdar}},
  \bibinfo{author}{\bibfnamefont{O.}~\bibnamefont{Narayan}}, \bibnamefont{and}
  \bibinfo{author}{\bibfnamefont{T.~A.} \bibnamefont{Witten}},
  \bibinfo{journal}{Phys. Rev. E} \textbf{\bibinfo{volume}{53}},
  \bibinfo{pages}{4673} (\bibinfo{year}{1996}).

\bibitem[{\citenamefont{Liu and Nagel}(2001)}]{liu01}
\bibinfo{editor}{\bibfnamefont{A.~J.} \bibnamefont{Liu}} \bibnamefont{and}
  \bibinfo{editor}{\bibfnamefont{S.~R.} \bibnamefont{Nagel}}, eds.,
  \emph{\bibinfo{title}{Jamming and Rheology}} (\bibinfo{publisher}{Taylor \&
  Francis}, \bibinfo{address}{London}, \bibinfo{year}{2001}).

\bibitem[{\citenamefont{Voss}(1985)}]{voss85}
\bibinfo{author}{\bibfnamefont{R.~F.} \bibnamefont{Voss}}
  (\bibinfo{publisher}{Springer-Verlag}, \bibinfo{address}{Berlin},
  \bibinfo{year}{1985}), p. \bibinfo{pages}{805}.

\bibitem[{\citenamefont{Molinari and Ortiz}(2002)}]{molinari00c}
\bibinfo{author}{\bibfnamefont{J.~F.} \bibnamefont{Molinari}} \bibnamefont{and}
  \bibinfo{author}{\bibfnamefont{M.}~\bibnamefont{Ortiz}},
  \bibinfo{journal}{Int. J. Numer. Methods in Eng.}
  \textbf{\bibinfo{volume}{53}}, \bibinfo{pages}{1101} (\bibinfo{year}{2002}),
  \bibinfo{note}{for a description of adaptive refinement schemes in large
  deformation problems see,}.

\bibitem[{\citenamefont{Belytschko and Hughes}(1983)}]{belytschko83}
\bibinfo{author}{\bibfnamefont{T.}~\bibnamefont{Belytschko}} \bibnamefont{and}
  \bibinfo{author}{\bibfnamefont{T.~J.~R.} \bibnamefont{Hughes}},
  \emph{\bibinfo{title}{Computational Methods for Transient Analysis}}
  (\bibinfo{publisher}{North-Holland}, \bibinfo{address}{Amsterdam},
  \bibinfo{year}{1983}).

\bibitem[{\citenamefont{Hughes}(1987)}]{hughes87}
\bibinfo{author}{\bibfnamefont{T.~J.~R.} \bibnamefont{Hughes}},
  \emph{\bibinfo{title}{The finite element method: linear static and dynamic
  finite element analysis}} (\bibinfo{publisher}{Prentice-Hall},
  \bibinfo{address}{Englewood Cliffs}, \bibinfo{year}{1987}).

\bibitem[{\citenamefont{Molinari et~al.}(2001)\citenamefont{Molinari, Ortiz,
  Radovitzky, and Repetto}}]{molinari01}
\bibinfo{author}{\bibfnamefont{J.~F.} \bibnamefont{Molinari}},
  \bibinfo{author}{\bibfnamefont{M.}~\bibnamefont{Ortiz}},
  \bibinfo{author}{\bibfnamefont{R.}~\bibnamefont{Radovitzky}},
  \bibnamefont{and} \bibinfo{author}{\bibfnamefont{E.~A.}
  \bibnamefont{Repetto}}, \bibinfo{journal}{Engineering Computations}
  \textbf{\bibinfo{volume}{18}}, \bibinfo{pages}{592} (\bibinfo{year}{2001}).

\bibitem[{Har()}]{Harrison}
\bibinfo{note}{J. A. Harrison, S. Hyun, J. F. Molinari and M. O. Robbins,
  unpublished.}

\bibitem[{foo()}]{footnext}
\bibinfo{note}{We also considered connections between next-nearest neighbors.
  This increases the cluster sizes, leading to a distribution much like that
  for $h_c=0.025$ (Fig. 11). However, it does not change the fundamental
  difference between clusters produced by the full calculation and those from
  the rigid overlap model. Note that including next-nearest neighbor
  connections has no noticeable affect on $P(a_c)$ in the rigid overlap model.}

\bibitem[{\citenamefont{Kondev and Henley}(1995)}]{kondev95}
\bibinfo{author}{\bibfnamefont{J.}~\bibnamefont{Kondev}} \bibnamefont{and}
  \bibinfo{author}{\bibfnamefont{C.~L.} \bibnamefont{Henley}},
  \bibinfo{journal}{Phys. Rev. Lett.} \textbf{\bibinfo{volume}{74}},
  \bibinfo{pages}{4580} (\bibinfo{year}{1995}).

\bibitem[{\citenamefont{O'Hern et~al.}(2001)\citenamefont{O'Hern, Langer, Liu,
  and Nagel}}]{ohern01}
\bibinfo{author}{\bibfnamefont{C.~S.} \bibnamefont{O'Hern}},
  \bibinfo{author}{\bibfnamefont{S.~A.} \bibnamefont{Langer}},
  \bibinfo{author}{\bibfnamefont{A.~J.} \bibnamefont{Liu}}, \bibnamefont{and}
  \bibinfo{author}{\bibfnamefont{S.~R.} \bibnamefont{Nagel}},
  \bibinfo{journal}{Phys. Rev. Lett.} \textbf{\bibinfo{volume}{86}},
  \bibinfo{pages}{111} (\bibinfo{year}{2001}).

\bibitem[{\citenamefont{Rottler and Robbins}(2002)}]{rottler02}
\bibinfo{author}{\bibfnamefont{J.}~\bibnamefont{Rottler}} \bibnamefont{and}
  \bibinfo{author}{\bibfnamefont{M.~O.} \bibnamefont{Robbins}},
  \bibinfo{journal}{Phys. Rev. Lett.} \textbf{\bibinfo{volume}{89}},
  \bibinfo{pages}{195501} (\bibinfo{year}{2002}).

\end{thebibliography}
\end{document}